%% file: proc.tex
\documentclass[cits]{PoS}
\usepackage{graphics}
\usepackage{epsfig}
\usepackage{amsmath,amssymb}

\title{Chiral low-energy constants from lattice QCD}
\ShortTitle{Chiral low-energy constants from lattice QCD}
\author{Silvia Necco\thanks{Work partially supported 
by EC Sixth Framework Program under the contract MRTN-CT-2006-035482
(FLAVIAnet), by the Ministerio de Ciencia e Innovaci\'on under Grant  No. FPA2007-60323 and by CPAN (Grant No. CSD2007-00042).
 Preprint numbers: IFIC/08-62, CERN-PH-TH-2009-015.
}\\
CERN, Physics Departement, 1211 Geneva 23, Switzerland\\
        E-mail: \email{Silvia.Necco@cern.ch}}

\abstract{Different strategies for the computation of QCD low-energy
  couplings by matching lattice results with the chiral effective theory
  are reviewed. After recalling some relevant predictions from the
  effective theory, the current status of leading order ($\Sigma,F,
  \Sigma_0,F_0$) and next-to-leading order ($l_i, L_i$) low-energy
  constants is summarized, focusing on recent results obtained with
  $N_f=2$ and $N_f=2+1$ lattice simulations.}

\FullConference{8th Conference Quark Confinement and the Hadron Spectrum \\
                 September 1-6 2008\\
                 Mainz, Germany}
\begin{document}
\input{intro}

\input{pred}
\input{alter}

\input{summary}


\input proc.bbl
\end{document}

%% file: intro.tex
\section{Introduction}
The dynamics of QCD at low momenta can be described in terms of an
effective theory, which encodes the spontaneous breaking of
chiral symmetry
$$
{\rm SU}(N_f)_L\times {\rm SU}(N_f)_R\rightarrow   {\rm SU}(N_f)_V.
$$
$N_f$ is the number of light quark flavours; we will consider both
cases $N_f=2,3$.
The effective Lagrangian in Euclidean space can be expanded in powers of
momenta as \cite{Weinberg:1978kz,Gasser:1983yg,Gasser:1984gg}
\begin{equation}
\mathcal{L}_\chi=\mathcal{L}_\chi^{(2)}+\mathcal{L}_\chi^{(4)}+\cdots,
\end{equation}
where 
\begin{eqnarray}
\mathcal{L}_{\chi}^{(2)} & = & \frac{F^2}{4}{\rm Tr} \left[\partial_{\mu}U^\dagger \partial_{\mu}U\right] -\frac{\Sigma}{2}
{\rm Tr} \left[\mathcal{M}U+ U^\dagger \mathcal{M}^\dagger       \right],\label{lagr}\\
\mathcal{L}_{\chi}^{(4)} & = & \sum_{i}{C}_i\mathcal{O}_i.
\end{eqnarray}
$U\in {\rm SU}(N_{f})$ represents the pseudo Nambu-Goldstone bosons
degrees of freedom, and $\mathcal{M}$ is the $(N_{ f} \times N_{  f})$ quark mass matrix.
The low-energy dynamics is parametrised by the so-called
Low-Energy-Couplings (LECs) of the effective theory, which are not
determined by symmetries. 
At leading order, the chiral Lagrangian involves two LECs, namely the
pseudoscalar decay constant $F$ and
the chiral condensate $\Sigma$ (or equivalently $B=\Sigma/F^2$) \footnote{In the following $F$,
    $\Sigma$ and $B$ will refer to the case $N_{f}=2$; for $N_{
      f}=3$ the notation $F_0$, $\Sigma_0$, $B_0$  will be used. }.
At next-to-leading order,
$\mathcal{L}_\chi^{(4)}$ contains 10 terms for the case ${ N_f}=2$
and 12 for ${N_f}=3$, with corresponding couplings
\begin{eqnarray}
\{{C}_i&\rightarrow& l_{i=1..7},h_{i=1..3}\;\;\;({N_f}=2),\\
\{{C}_i&\rightarrow& L_{i=1..10},H_{i=1..2}\;\;\;({N_f}=3).
\end{eqnarray}
Once the LECs are fixed, chiral
perturbation theory becomes a predictive framework 
and hence a powerful
tool for understanding the phenomenology of QCD at low dynamics. 
A review of recent applications has been presented at this conference by
G. Ecker \cite{Ecker:2008fv}. Those couplings must be ideally computed from''first
principles'': from this point of view lattice QCD is a promising approach, since -once the physical
point has been reached (i.e. continuum, infinite volume limit and physical light quark masses)-it does not introduce
any model dependence.

For some decades lattice simulations have been performed far from the
chiral limit and introducing simplifications such as quenching.
Chiral effective theory has been then extensively used
to understand the results of lattice QCD, in particular to guide
extrapolations to the chiral limit and to estimate the volume dependence 
of physical observables computed on the lattice. 
In the past years lattice computations experienced an important progress,
due to an interplay of new theoretical developments, algorithmic improvements and increasing powerful computing resources. 
The effect of this progress is that now lattice unquenched simulations
with $N_{ f}=2$ and $N_{f}=2+1$ light dynamical flavours are approaching domains where a reliable
matching with the chiral effective theory can be performed.
\emph{Reliable} in this case means that light quark masses are as
close as possible to the physical value, and that different sources of systematic errors can be at
least in principle taken under control, namely finite-volume effects, lattice
artifacts and renormalisation.

%% file: pred.tex
\section{Low-Energy couplings from quark-mass dependence of  pseudoscalar
  masses and decay constants}\label{sec2}
The LECs can be extracted for instance by studying the
quark-mass dependence of given observables in lattice QCD and match it with the
predictions of the chiral effective theory.
In this section I will summarize some of these predictions, in
particular for the pseudoscalar decay constants and masses.
The up and down quarks are clearly sufficiently light to
formulate a SU(2) $\times$ SU(2) chiral effective theory where pions are
treated as pseudo Nambu-Goldstone bosons and Kaons and etas are
integrated out. In this theory, the LECS will depend on the higher
energy scales $m_s,m_c,m_b,\Lambda_{\rm QCD}$. 
Alternatively, one can decide to treat also the strange quark as
light and formulate a  SU(3) $\times$ SU(3) effective theory, where
also Kaons and etas are treated as pseudo Nambu-Goldstone bosons and
the LECs will not depend on $m_s$ anymore.

One of the simplest but deepest predictions which can be formulated at
low energy is the so-called Gell-Mann-Oakes-Renner relation \cite{GellMann:1968rz}, which
states that the square of the pion mass is linearly proportional to
the light quark mass:
\begin{equation}
{M_\pi^2}=M^2=2\hat{m}B,
\end{equation}
where $\hat{m}=(m_u+m_d)/2$. This corresponds to the
leading order prediction in the effective theory. Higher order
corrections can be computed systematically.
In particular for the SU(2) $\times$ SU(2) theory one finds the NLO
expressions \cite{Gasser:1983yg}
\begin{eqnarray}
M_\pi^2 &=& M^2+\frac{M^4}{32\pi^2F^2}\ln\left(\frac{M^2}{\Lambda_3^2}\right),\\
F_\pi & = & F-\frac{M^2}{16\pi^2F^2}\ln\left(\frac{M^2}{\Lambda_4^2}\right),
\end{eqnarray} 
where $\Lambda_{3,4}$ are related to the scale-independent
NLO LECs $\bar{l}_{3,4}$ by the relation
\begin{equation}
\bar{l}_{3,4}\equiv \ln\left(\frac{\Lambda_{3,4}^2}{M^2}
\right)_{M=139.6\;{\rm MeV}}.
\end{equation}
By matching the quark-mass dependence of $F_\pi$ and $M_\pi$ with
lattice results, it is hence possible to extract the LO LECs $F$ and
$\Sigma$ (or equivalently $B$) and the NLO couplings $\bar{l}_3,\bar{l}_4$. \\
In this theory there is the possibility to treat Kaons as external matter fields coupled
to pions, in analogy to what is done in the heavy-light meson chiral
perturbation theory. The quark mass dependence of Kaon
quantities can then be worked out explicitly in this framework \cite{Roessl:1999iu}.

In the SU(3) $\times$ SU(3) theory, the quark mass dependence of
pion and Kaon masses and decay constants is given by \cite{Gasser:1984gg}
\begin{eqnarray}
M_\pi^2 & =& 2\hat{m}{B_0}\left\{1+{\mu_\pi}-\frac{1}{3}{\mu_\eta}+2\hat{m}{K_3} +{K_4}   \right\},   \\
M_K^2 & =&  (\hat{m}+m_s){B_0}
\left\{1+\frac{2}{3}{\mu_\eta}+(\hat{m}+m_s){K_3}+{K_4}   \right\},  \\
F_\pi & =&
{F_0}\left\{1-2{\mu_\pi}-{\mu_K}+2\hat{m}
{K_6} +{K_7}   \right\},   \\
F_K & =&  {F_0}
\left\{1-\frac{3}{4}{\mu_\pi}-\frac{3}{2}{\mu_K}-\frac{3}{4}\mu_\eta+
(\hat{m}+m_s){K_6}+{K_7}   \right\},
\end{eqnarray}
with
$$
\mu_P=\frac{M_P^2}{32\pi^2F_0^2}\ln\left(\frac{M_P^2}{\mu^2}   \right).
$$
The $K_i$ contain the NLO LECs:
\begin{eqnarray}
{K_3}=\frac{8{B_0}}{{F_0}^2}({2L_8-L_5}),&\;\;\;\;&
{K_4}=(m_u+m_d+m_s)\frac{16{B_0}}{{F_0}^2}({2L_6-L_4}),\\
{K_6}=\frac{4{B_0}}{{F_0}^2}({L_5}),&\;\;\;\;&{K_7}=(m_u+m_d+m_s)\frac{8{B_0}}{{F_0}^2}({L_4}).
\end{eqnarray}
In this case, the matching with lattice results allows the extraction
of $F_0,\Sigma_0(B_0), L_4,L_5,\;(2L_8-L_5),\;(2L_6-L_4)$. Moreover, the $m_s$-dependence of the SU(2) $\times$ SU(2) LECs can be
explicitly worked out \cite{Gasser:1984gg}
\begin{eqnarray}
{F} &=& {F_0}\left\{1+8\frac{\overline{M}_K^2}{F_0^2}L_4-\overline{\mu}_K+O(m_s^2)
\right\},\\
{\Sigma} &=&
{\Sigma_0}\left\{1+32\frac{\overline{M}_K^2}{F_0^2}L_6-2\overline{\mu}_K-\overline{\mu}_\eta+O(m_s^2)
\right\},
\end{eqnarray}
where $\overline{M}_K$ is the Kaon mass in the limit
$m_u,m_d\rightarrow 0$, $\overline{M}_K^2=m_sB_0$.
In the large-$N_c$ limit one finds $\Sigma/\Sigma_0,F/F_0,
B/B_0\rightarrow 1$, which corresponds to the so-called
Okubo-Zweig-Iizuka (OZI) rule.
Paramagnetic inequalities can be invoked to predict the sign of the
deviations from the OZI rule, $(F/F_0-1)>0, (\Sigma/\Sigma_0-1)>0$ \cite{DescotesGenon:1999uh}.

Lattice simulations are performed on finite boxes with volume $V=L^3 T$, and finite-volume effects
may play a relevant role. An important point is that chiral perturbation theory can be used
also to infer information about finite-size scaling of physical
observables \cite{Gasser:1986vb,Gasser:1987ah,Gasser:1987zq}.
In the asymptotic region $M_\pi L\gg 1$, the dominant effect is of the
form $e^{-M_\pi L}$. The finite-size effects for the pion mass have been
computed at two loops \cite{Colangelo:2006mp}; alternatively, 
resummations of asymptotic L\"uscher formulas 
\cite{Luscher:1985dn} for pseudoscalar masses and decay constants have
been investigated \cite{Colangelo:2005gd}.

Lattice calculations are often performed in so-called
\emph{partially quenched} setup, where $m_{sea}\neq m_{val}$. 
It is possible to extend the effective theory formalism to this case and
obtain predictions from \emph{partially quenched chiral perturbation theory}
\cite{Sharpe:2000bc} which can be matched with the lattice results.

Finally, lattice simulations are performed at small but finite lattice
spacing $a$; ideally
one should first perform a continuum extrapolation of lattice results and then match with
the chiral effective theory. In alternative, one can formulate the
chiral effective theory at finite lattice spacing by adopting the
Symanzik formalism \cite{Symanzik:1983dc,Symanzik:1983gh}. The price to pay is the appearing of extra
couplings in the effective theory which have to be determined by the
matching. Notice that for particular discretizations, where flavour symmetry is broken at finite lattice spacing,
 for example for the so-called
staggered fermions with fourth root prescription, the limits $a\rightarrow 0$ and $m\rightarrow 0$
have to be taken simultaneously. Lattice results must be then
matched with the so-called \emph{staggered chiral perturbation theory} \cite{Lee:1999zxa}.

In several recent lattice studies with ${N_f}=2$ dynamical
fermions the quark mass dependence of $M_\pi$ and $F_\pi$ has been
matched with SU(2) $\times$ SU(2) chiral perturbation theory in order
to extract $F$, $\Sigma$, $\overline{l}_3$, $\overline{l}_4$.
The relevant parameters of these simulations are summarised in Table
\ref{nf2tab}: the discretization adopted, the lattice spacing (in
parenthesis the quantity that has been used to fix the scale), the
pseudoscalar meson masses and the volume in terms of the quantity $M_\pi L$.\\
One of the main outcomes of these studies is that $M_\pi^2$ is nearly a linear
function of $\hat{m}$, as predicted by the GMOR relation, up to relatively large masses (of the order of
$m_s/2$). 
A further common conclusion is that for $M_\pi\lesssim 450$ MeV lattice
results seem compatible with prediction of the chiral effective
theory at NLO. 
CERN-TOV and JLQCD pointed out however that for $F_\pi$ NNLO effects may
be significant and this could affect the systematic uncertainty on the
determination of $\bar{l}_4$.

Several collaborations are performing lattice calculations with 2+1
dynamical flavours, i.e. with two light (degenerate) quarks and
one heavier quark (which is usually fixed at the physical strange
mass $m_s$). In analogy to the $N_f=2$ case, we report the relevant
simulation parameters in Table \ref{nf3tab} \footnote{Another collaboration has recently published results for hadron masses obtained with $2+1$ lattice simulations \protect\cite{Durr:2008zz}.}. The MILC data have been
analysed using SU(3) $\times$ SU(3) (partially quenched) rooted staggered chiral perturbation theory
including analytic NNLO and NNNLO terms. The NPLQCD collaboration adopted Domain Wall valence quarks and staggered sea quarks, and fitted the data using continuum chiral perturbation theory at NLO and (partial) NNLO.
RBC/UKQCD adopted a partially quenched setup, while PACS-CS used
only unitary points. 
These two collaborations matched their results for pseudoscalar masses and
decay constants with SU(3) $\times$ SU(3) effective theory at NLO and observed poor convergence of the perturbative series around the physical
strange quark mass. On the other hand, the NLO SU(2) $\times$ SU(2) analysis seems to
yield more reliable results.\\
The results obtained for the LECs are all summarized in the Tables \ref{LOrestab}, \ref{LOres1tab}, \ref{NLOres1tab}, \ref{NLOres2tab}.

\begin{table}
\begin{tabular}{ccccc}
\hline
& & & & \\[-0.2cm]
Collaboration & Dirac operator &   $a$ (fm) & $M_\pi$ (MeV) & $M_{\pi,min}L$ \\
\hline
& & & & \\[-0.2cm]
CERN-TOV \protect\cite{DelDebbio:2006cn}  & Wilson +$O(a)$ impr. & 0.052-0.072
($M_K$)  & $\gtrsim$ 380  &  3.2 -3.6 \\
ETM \protect\cite{Boucaud:2007uk, Dimopoulos:2008sy} & Wilson TM & 0.065-0.1 ($F_\pi$)& $\gtrsim$ 265 & 3.2-3.6\\
JLQCD/TWQCD \protect\cite{Noaki:2008iy} & Neuberger  & 0.12 ($r_0$) & $\gtrsim$ 290 & 2.9\\ 
\hline
\end{tabular}
\caption{Simulation parameters of recent $N_f=2$ computations.}\label{nf2tab}
\end{table}

\begin{table}
\begin{tabular}{ccccc} 
\hline
& & & & \\[-0.3cm]
Collaboration &  Dirac operator &  $a$ (fm)   & $M_{\pi}$ (MeV) & $M_{\pi,{\rm min}}L$
\\
\hline
& & & &\\[-0.2cm]
MILC \protect\cite{Aubin:2004fs,Bernard:2007ps} & Staggered &  0.06-0.18[$F_\pi$]   & $\gtrsim$ 240 & 4\\
NPLQCD \protect\cite{Beane:2006kx} & Domain Wall + Staggered & 0.13 [$r_0$] & $\gtrsim$ 290 & 3.7\\
RBC/UKQCD \protect\cite{Allton:2008pn} & Domain Wall & 0.11 [$M_\Omega$]  & $\gtrsim$ 330 & 4.6\\
PACS-CS \protect\cite{Aoki:2008sm} & Wilson + $O(a)$ impr.  & 0.09 [$M_{\Omega}$]   & $\gtrsim$ 160 & 2.3 \\[0.1cm]
\hline
\end{tabular}
\caption{Simulation parameters of recent $N_f=2+1$ computations. $r_0$ is the so-called Sommer scale \protect\cite{Sommer:1993ce}. }\label{nf3tab}
\end{table}

%% file: alter.tex
\section{Other approaches to determine LECs on the lattice}
\subsection{LECs from $\epsilon$-regime simulations}
On a finite volume $V=L^3T$ with $L\gg 1/\Lambda_{\rm QCD}$, different
chiral regimes can be distinguished.
Approaching the chiral limit by keeping $M_\pi L\gg 1 $ (like already mentioned in Section \ref{sec2}) defines the
so-called $p$-regime, where finite-volume effects are exponentially
suppressed, while mass-effects are dominant.
Alternatively, one can approach the chiral limit by keeping
$\mu=m\Sigma V\lesssim O(1)$; this defines the so-called
$\epsilon$-regime \cite{Gasser:1986vb,Gasser:1987ah}, where the
Compton wavelength associated to the Nambu-Goldstone bosons is larger
than the linear extent of the box, $M_\pi L<1$. In this case the power
counting is reorganised such that mass effects are suppressed, while
finite-volume effects are enhanced and become polynomial in
$L^{-2}$. One of the consequences of the rearrangements is that, at a
given order in the effective theory, less LECs appear with respect to
the $p$-regime, and the predictions are less ``contaminated'' by higher order unknown couplings.
For instance, the NLO predictions of given
correlation functions contain only the LO couplings: 
$F$, $\Sigma$ ($F_0$,$\Sigma_0$) can then be extracted by matching the finite-size scaling of correlators
computed on the lattice with the predictions of the
effective theory.\\
As an example, the pseudoscalar and axial correlators at NLO can be
written as (for $N_f=2$) \cite{Hansen:1990un}
\begin{eqnarray}
C_P(t) & = & \frac{1}{L^3}\int d^3 \vec{x}\langle P(x)P(0) \rangle =
\Sigma^2\left[a_P+\frac{T}{F^2L^3}b_Ph_1\left(\frac{t}{T}  \right)       \right],  \\
C_A(t) & = & \frac{1}{L^3}\int d^3 \vec{x}\langle A_0(x)A_0(0)
\rangle=\frac{F^2}{V}\left[a_A+ \frac{T}{F^2L^3}b_Ah_1\left(\frac{t}{T}  \right)       \right],
\end{eqnarray}
where $h_1(\tau)=\frac{1}{2}\left[\left(\tau-\frac{1}{2}
  \right)^2-\frac{1}{12}  \right]$, and $a_P,b_P,a_A,b_A$ are dimensionless
functions of $\mu$, $L$, $T$.\\
Furthermore, in the $\epsilon$-regime topology plays a relevant role
\cite{Leutwyler:1992yt}: observables may be defined at fixed value of
the topological charge, and the dependence on this charge should be
also reproduced by the chiral effective theory.\\
This strategy has been applied recently in two studies.
JLQCD \cite{Fukaya:2007pn}  computed mesonic correlation functions at fixed topology in the $\epsilon$-regime using the Neuberger Dirac operator with $N_f=2$ dynamical quarks. A. Hasenfratz and collaborators \cite{Hasenfratz:2008ce} adopted improved Wilson fermions combined with the reweighting technique.
The results obtained for $F$ and $\Sigma$ are reported in the summary Table \ref{LOrestab} (8,9).
The two-point functions computed in \cite{Hasenfratz:2008ce} have been 
recently reanalysed in \cite{Bar:2008th} including the leading $O(a^2)$ correction inferred via the so-called \emph{Wilson chiral perturbation theory} applied to the $\epsilon$-regime \cite{Bar:2008th,Shindler:2008ri}, finding small corrections with respect to the continuum results.

Two-point mesonic functions have been computed in the effective theory also in the case of non-degenerate quark masses, and in particular in the case where some quarks are in the $p$-regime and others in the $\epsilon$-regime \cite{Damgaard:2007ep,Bernardoni:2007hi,Bernardoni:2008ei}. This may offer new possibilities to extract the LECs from lattice data.
\subsection{LECs from eigenvalues distribution}
At LO in the $\epsilon$-expansion, the partition function is equivalent to the one of a chiral Random Matrix Theory (RMT) \cite{Shuryak:1992pi,Verbaarschot:1993pm,Verbaarschot:1994qf,Verbaarschot:2000dy}; it follows that RMT reproduces the same microscopic spectral density 
$\rho_S(\zeta,\mu)$ of the chiral effective theory, in terms of two dimensionless variables $\zeta=\lambda\Sigma V$ and $\mu=m\Sigma V$, where $\lambda$ represents the eigenvalues of the Dirac operator. Moreover, one can extract the probability distributions of single eigenvalues \cite{Nishigaki:1998is,
Damgaard:2000ah,Basile:2007ki}; hence it is possible to match the QCD low-lying spectrum of the Dirac operator $\langle \lambda_k\rangle^{\rm QCD}$
with the expectation values $\langle \zeta_k\rangle^{\rm RMT}$ (where $k=1,2..$ labels the eigenvalues) in order to extract the chiral condensate $\Sigma$.
This method has been applied by many authors;
the results obtained by the  JLQCD/TWQCD collaboration \cite{Fukaya:2007yv}  an by DeGrand and collaborators \cite{DeGrand:2006nv} are reported in Table \ref{LOrestab} (10,11).

This framework can be extended such that the spectrum of the Dirac operator is sensitive also to the pseudoscalar decay constant $F$ at LO in the effective theory \cite{Akemann:2006ru}. This strategy has been adopted  by DeGrand and collaborators \cite{DeGrand:2007tm}, yielding the results reported in Table \ref{LOrestab} (12).

In another recent work \cite{Giusti:2008vb} it is shown that 
the quark condensate can be extracted from suitable (renormalizable) spectral observables defined in the $p$-regime, for instance the number of Dirac operator modes contained in a given interval. The method has been used on Wilson lattice QCD and the result for the condensate is also given in Table  \ref{LOrestab} (6).
In correlation to this, an interesting development from the side of the effective theory has been investigated in \cite{Damgaard:2008zs}, where the quark condensate and the spectral density of the Dirac operator have been computed by using a technique which is able to smoothly connect $p$- and $\epsilon$-regimes.
\subsection{LECs from form factors}
The pion electromagnetic form factor, defined as
\begin{equation}
\langle \pi^+(p')|V_\mu|\pi^+(p)\rangle=(p+p')_\mu F^{\pi\pi}_V(q^2)
\end{equation}
with $q^2=-Q^2=(p-p')^2$, can be matched with the predictions of the chiral effective theory. In particular, at NLO it turns out to be related to the LEC $l_6$ for the SU(2) $\times$ SU(2) theory \cite{Gasser:1983yg} and to $L_9$ in the SU(3) $\times$ SU(3) case \cite{Gasser:1984ux}.  
The pion electromagnetic form factor has been computed on the lattice and compared with chiral effective theory by the RBC/UKQCD collaboration \cite{Boyle:2008yd} and by the ETM collaboration \cite{Frezzotti:2008dr}, using the lattice parameters already given in Tables \ref{nf2tab}, \ref{nf3tab}. 
The latter performed a NNLO chiral fit, which -by using as input the experimental value of the scalar pion radius- allowed the extraction of $F, \Sigma,\bar{l}_1, \bar{l}_2, \bar{l}_3, \bar{l}_4, \bar{l}_6$. We report some of the results in Tables \ref{LOrestab},\ref{NLOres1tab}. \\
Moreover, an exploratory study of the scalar form factor has been presented by the JLQCD/TWQCD collaboration at the last lattice conference \cite{Kaneko:2008kx}.


%% file: summary.tex
\section{Summary and Conclusions}
In Tables \ref{LOrestab}, \ref{LOres1tab}, \ref{NLOres1tab}, \ref{NLOres2tab} we summarize the results for the low-energy couplings obtained through lattice simulations.

In particular, Table \ref{LOrestab} collects the results for the LO LECs, both from $p$-regime and $\epsilon$-regime studies. The first error is statistical, while the following uncertainties (if present) are systematic. Although errors are still quite sizeable, the general agreement among the different determinations is reassuring. 
PACS-CS results for the condensate  $\Sigma$ point towards larger values with respect to other collaboration: notice however that in that case perturbative renormalization has been used and systematic errors may be important.
Table \ref{LOres1tab} shows the ratios $F/F_0$, $\Sigma/\Sigma_0$; large deviations from 1 would indicate strong violations of the OZI rule. More precise determinations are needed for these observables in order to draw a definitive conclusion.

Table \ref{NLOres1tab} summarizes the results for the SU(2) $\times$ SU(2) NLO constants $\bar{l}_3$, $\bar{l}_4$, $\bar{l}_6$. For some of the lattice simulations with $N_f=2+1$, we report both values obtained through a direct SU(2) $\times$ SU(2) fit and by converting SU(3) $\times$ SU(3) into SU(2) $\times$ SU(2) LECs by means of chiral perturbation theory. Also in this case the good agreement is encouraging. Lattice data for $\bar{l}_3$, $\bar{l}_4$ agree also with phenomenological estimates \cite{Gasser:1983yg,Colangelo:2001df}. The phenomenological impact of the LECs $\bar{l}_3$, $\bar{l}_4$
 on the s-wave pion scattering length $a_0^I (I=0,2)$ has been recently discussed by H. Leutwyler \cite{Leutwyler:2008fi}.

Finally, Table \ref{NLOres2tab} shows the lattice results for  SU(3) $\times$ SU(3) NLO constants $L_4$, $L_5$ and the combinations $(2L_6-L_4)$, $(2L_8-L_5)$, at the scale $M_\rho=770$ MeV. These values can be compared with phenomenological estimates reported in \cite{Bijnens:2007yd}. In this case the situation is less clear, since it has been pointed out that SU(3) $\times$ SU(3) chiral fits may not be appropriate to describe lattice results obtained with $N_f=2+1$.
Notice that in order to extract SU(3) $\times$ SU(3) couplings, lattice simulations with $N_f=3$ light flavours would be more appropriate.

To conclude, important progress occurred in lattice QCD simulation in the past years. Unquenched computations are reaching
pion masses as light as 200 MeV, lattice spacings around $a\sim 0.07$ fm and spatial extends larger than 4 fm, moving towards the ``physical point''.
 This will allow a precise matching with the chiral effective theory and a deep understanding of low-energy properties of strong interactions. 
By decreasing systematic uncertainties,
systematic errors become a very important issue and have to be carefully studied, namely by controlling the continuum extrapolation, the finite-volume effects, the renormalisation and the uncertainty coming from higher orders in the chiral effective theory. 
LECs obtained with independent methods and different discretizations tend to point at uniform results within the errors, giving encouraging perspectives for the future.

\begin{table}
{\footnotesize{
\begin{tabular}{llllll}
\hline
& & & & & \\[-0.2cm]
Collaboration & $N_f$  & $F$ (MeV) & $F_0$ (MeV) &
$\Sigma^{1/3}$ (MeV)  & $\Sigma_0^{1/3}$ (MeV)\\
& & & & & \\[-0.2cm]
\hline
& & & & & \\[-0.2cm]
(1) {\footnotesize{ETM}}  \protect\cite{Boucaud:2007uk,Dimopoulos:2008sy} & 2   &  86.03(5) & & 267(2)(9)(4)
& \\
(2) {\footnotesize{JLQCD/TWQCD}} \protect\cite{Noaki:2008iy}& 2 &
79.0(2.5)(0.7)$\binom{+4.2}{-0.0}$  & &
235.7(5.0)(2.0)$\binom{+12.7}{-0.0}$ & \\
(3) {\footnotesize{MILC}} \protect\cite{Bernard:2007ps} & 2+1 &  & &$278(1)\binom{+2}{-3}(5)$  & $242(9)\binom{+5}{-17}(4)$\\
(4) {\footnotesize{RBC/UKQCD}} \protect\cite{Allton:2008pn} & 2+1 & 81.2(2.9)(5.7) & & 255(8)(8)(13)  & \\
(5) {\footnotesize{PACS-CS}} \protect\cite{Aoki:2008sm} & 2+1  & 90.3(3.6) & 83.8(6.4) &
309(7)  & 290(15) \\
(6) {\footnotesize{Giusti, L\"uscher}}\protect\cite{Giusti:2008vb} & 2 &
&  & 276(3)(4)(5)   &\\
(7)  {\footnotesize{ETM}}  \protect\cite{Frezzotti:2008dr} & 2   &  86.6(4)(7) & & 264(2)(5)
& \\
& & & & & \\[-0.2cm]
\hline
& & & & & \\[-0.2cm]
(8) {\footnotesize{JLQCD}}\protect\cite{Fukaya:2007pn} & 2 
& 87.3(5.6)   &  & 239.8(4.0)  & \\
(8) {\footnotesize{A. Hasenfratz \emph{et
      al}}}\protect\cite{Hasenfratz:2008ce} & 2 & 90(4)   &  & 248(6)  & \\
(10) {\footnotesize{JLQCD}}\protect\cite{Fukaya:2007yv} & 2 
&   &  & 251(7)(11)  & \\
(11)  {\footnotesize{DeGrand \emph{et al}}}\protect\cite{DeGrand:2006nv} & 2 
&  &  & 282(10)  & \\                            
(12) {\footnotesize{DeGrand, Schaefer}}\protect\cite{DeGrand:2007tm} & 2 
&  84(5) &  & 234(4)  & \\
& & & & & \\[-0.2cm]
\hline
& & & & & \\[-0.2cm]
Colangelo, D\"urr \protect\cite{Colangelo:2003hf} &  phen. &  86.2(5)& & &\\
\hline
\end{tabular}}}
\caption{Summary of lattice results for LO couplings. The condensate $\Sigma$ is evaluated in the $\overline{\rm MS}$  scheme at the scale $\mu=2$ GeV.
The upper part (1-7) collects results obtained by matching lattice QCD with the chiral effective theory in the $p$-regime, while the bottom part (8-12) summarizes results obtained in the $\epsilon$-regime.  
The last row is the estimation from a phenomenological analysis.
}\label{LOrestab}
\end{table}

\begin{table}
\begin{tabular}{lll}
\hline
& & \\[-0.2cm]
Collaboration & $F/F_0$ & $\Sigma/\Sigma_0$ \\
\hline
& & \\[-0.3cm]
{\footnotesize{MILC}} \protect\cite{Bernard:2007ps} & 1.15(5)$\binom{+13}{-3}$ & 1.52(17)$\binom{+38}{-15} $\\
{\footnotesize{RBC/UKQCD}} \protect\cite{Allton:2008pn}  & 1.229(59) & 1.55(21)\\
 {\footnotesize{PACS-CS}} \protect\cite{Aoki:2008sm} &  1.065(58)& 1.205(14
)\\
& & \\[-0.4cm]
\hline
\end{tabular}
\caption{Summary of results for the ratios $F/F_0$ and  $\Sigma/\Sigma_0$ obtained in recent $N_f=2+1$ lattice simulations. }\label{LOres1tab}
\end{table}


\begin{table}
{\footnotesize{
\begin{tabular}{lllllll}
\hline
& & & & & &\\[-0.4cm]
Collaboration & $N_f$ & $\bar{l}_3$ (SU(2)) & $\bar{l}_4$ (SU(2)) & $\bar{l}_3$ (SU(3)) &   $\bar{l}_4$ (SU(3))  & $\bar{l}_6$ (SU(2)) \\
\hline
 {\footnotesize{CERN-TOV}} \protect\cite{DelDebbio:2006cn} & 2 & 3.0(5)(1) &&& &\\
{\footnotesize{ETM}}  \protect\cite{Boucaud:2007uk,Dimopoulos:2008sy} & 2 & 3.42(8)(10)(27)
& 4.59(4)(2)(13) & &  &\\ 
{\footnotesize{ETM}}  \protect\cite{Frezzotti:2008dr} & 2 & 3.2(4)(2)
& 4.4(1)(1) & & & 14.9(6)(7)\\ 
 {\footnotesize{JLQCD/TWQCD}} \protect\cite{Noaki:2008iy} & 2 & 3.44(57)$\binom{+0}{-68}\binom{+32}{-0}$ &
4.14(26)$\binom{+49}{-0}\binom{+32}{-0}$ & & &\\[0.1cm]
 {\footnotesize{MILC}} \protect\cite{Bernard:2007ps} & 2+1  &  & &
$1.1(6)\binom{+1.0}{-1.5}$ & $4.4(4)\binom{+4}{-1}$ &\\
 {\footnotesize{RBC/UKQCD}} \protect\cite{Allton:2008pn} & 2+1 & 3.13(33)(24) & 4.43(14)(77) &
2.87(28) & 4.10(5)  &\\
 {\footnotesize{RBC/UKQCD}} \protect\cite{Boyle:2008yd} & 2+1 &   &   &&&   12.24(67)(71) \\
 {\footnotesize{PACS-CS}} \protect\cite{Aoki:2008sm} & 2+1 &  3.14(23) & 4.04(19) & 3.47(11) & 4.21(11) &\\ 
\hline
& & & & \\[-0.4cm]
{\footnotesize{Gasser, Leutwyler}} \protect\cite{Gasser:1983yg} & phen. & 2.9(2.4) & 4.3(9)&  &\\
{\footnotesize{Colangelo \emph{et al}}} \protect\cite{Colangelo:2001df} & phen.  &  & 4.4(2) &  &\\
& & & & \\[-0.4cm]
\hline
\end{tabular}}}
\caption{NLO couplings $\bar{l}_3$, $\bar{l}_4$, $\bar{l}_6$ of the SU(2) $\times$
  SU(2) chiral effective theory. The ``SU(2)'' label means that the
  LECs have been extracted directly from a SU(2) fit of lattice data,
  while the label ``SU(3)'' indicates the values obtained via the NLO relations 
   which convert SU(3) NLO couplings into the SU(2) ones.  The last rows refer to phenomenological estimates.}\label{NLOres1tab}
\end{table}


\begin{table}
\begin{tabular}{lllll}
\hline
& & & & \\[-0.4cm]
Collaboration & $L_4\cdot 10^3$ & $L_5\cdot 10^3$ & $(2L_6-L_4)\cdot 10^3$  &
$(2L_8-L_5)\cdot 10^3$ \\
\hline
& & & &\\[-0.3cm]
{\footnotesize{MILC}} \protect\cite{Bernard:2007ps}& $0.1(3)\binom{+3}{-1}$ & $1.4(2)\binom{+2}{-1}$ &
$0.3(1)\binom{+2}{-3}$ & 0.3(1)(1) \\
{\footnotesize{RBC/UKQCD}}  \protect\cite{Allton:2008pn} & 0.14(8)  & 0.87(10) & 0.00(4) & 0.24(4) \\
{\footnotesize{PACS-CS}}  \protect\cite{Aoki:2008sm}  & -0.06(10) & 1.45(7) & 0.10(2) & -0.21(3) \\
{\footnotesize{NPLQCD}}  \protect\cite{Beane:2006kx}  &  & 1.42(2)$\binom{+18}{-54} $&  &  \\
\hline
& & & &\\[-0.4cm]
{\footnotesize{Bijnens (lat 07)}} \protect\cite{Bijnens:2007yd}   &0  & 1.46 & 0& 0.54\\
& & & &\\[-0.4cm]
\hline
\end{tabular}
\caption{NLO SU(3) couplings at the scale $M_\rho=770$ MeV extracted
  from $N_f=2+1$ lattice simulations. In the last line the phenomenological estimate is reported.}\label{NLOres2tab}
\end{table}